\documentclass[%
 reprint,
superscriptaddress,
 amsmath,amssymb,
 aps,
]{revtex4-1}

\usepackage{booktabs}
\usepackage{multirow}

\usepackage{subcaption}

\usepackage{bm} 
\usepackage{graphicx}
\usepackage{dcolumn}
\usepackage{bm}
\usepackage{hyperref}

\begin{document}

\preprint{APS/123-QED}

\title{The Lensing Effect of Quantum-Corrected Black Hole and Parameter Constraints from EHT Observations}

\author{Lai Zhao}

\author{Meirong Tang}

\author{Zhaoyi Xu}%
\email{zyxu@gzu.edu.cn(Corresponding author)}
\affiliation{%
 College of Physics,Guizhou University,Guiyang,550025,China
}%


\begin{abstract}
The quantum-corrected black hole model demonstrates significant potential in the study of gravitational lensing effects. By incorporating quantum effects, this model addresses the singularity problem in classical black holes. In this paper, we investigate the impact of the quantum correction parameter on the lensing effect based on the quantum-corrected black hole model. Using the black holes $M87^*$ and $Sgr A^*$ as our subjects, we explore the influence of the quantum correction parameter on angular position, Einstein ring, and time delay. Additionally, we use data from the Event Horizon Telescope observations of black hole shadows to constrain the quantum correction parameter.
Our results indicate that the quantum correction parameter significantly affects the lensing coefficients $\bar{a}$ and $\bar{b}$, as well as the Einstein ring.
The position $\theta_{\infty}$ and brightness ratio $S$ of the relativistic image exhibit significant changes,with deviations on the order of magnitude of $\sim 1\mu as$ and $\sim 0.01\mu as$, respectively.
The impact of the quantum correction parameter on the time delay $\Delta T_{21}$ is particularly significant in the $M87^*$ black hole, with deviations reaching up to several tens of hours. Using observational data from the Event Horizon Telescope(EHT) of black hole shadows to constrain the quantum correction parameter, the constraint range under the $M87^*$ black hole is $0\le \frac{\alpha}{M^2}\le 1.4087$ and the constraint range under  the $Sgr A^*$ black hole is 
$0.9713\le \frac{\alpha}{M^2}\le 1.6715$ . Although the current resolution of the EHT limits the observation of subtle differences, future high-resolution telescopes are expected to further distinguish between the quantum-corrected black hole and the Schwarzschild black hole, providing new avenues for exploring quantum gravitational effects.

\begin{description}
\item[Keywords]
Quantum-Corrected Black Hole, Gravitational Lensing, Einstein Ring, Time Delay, Event Horizon Telescope
\end{description}
\end{abstract}

\maketitle


\section{\label{sec:level1}Introduction}
General relativity(GR) theoretically predicts the existence of black holes. In 2015-2016, LIGO's first detection of gravitational waves from the merger of binary black holes provided the first direct evidence of black holes in the universe, offering solid observational support for general relativity \cite{LIGOScientific:2016emj}. Moreover, GR has been extensively validated in other areas, such as through a series of tests in cosmology and pulsars \cite{Will:2014kxa,Ferreira:2019xrr,Clifton:2011jh}, as well as through observations by the EHT \cite{EventHorizonTelescope:2019dse,EventHorizonTelescope:2022wkp}. However, GR also has its limitations, particularly under certain conditions of matter and energy, where gravitational collapse inevitably leads to the formation of spacetime singularities. This phenomenon is encapsulated in the famous singularity theorems proposed by Hawking and Penrose \cite{Hawking:1970zqf,Penrose:1964wq}. Near these singularities, it is widely recognized that all physical measurements become divergent. To address these "singularities," Penrose proposed the cosmic censorship conjecture. Many scholars have tested this conjecture, as evidenced by numerous studies 
\cite{Ames:2023akp,Zhao:2024qzg,Meng:2024els,Zhao:2024lts,Tang:2023sig,Meng:2023vkg,Sadeghi:2023lkz,Zhao:2023vxq}. 
However, some researchers argue that considering quantum effects could avoid singularities \cite{Ashtekar:2006wn}. Within this theoretical framework, loop quantum gravity stands out as one of the primary candidates.

Loop quantum gravity(LQG) is a highly regarded quantum gravity theory, characterized by its background independence and non-perturbative nature \cite{Yang:2022btw,Ashtekar:2004eh,Han:2005km,Perez:2012wv}. As a result, it has garnered significant attention and has been extensively studied \cite{Perez:2012wv,Giesel:2011idc,Thiemann:2002nj,Ashtekar:2004eh}.
In response to the problem in GR where gravitational collapse inevitably leads to the formation of spacetime singularities, research in LQG theory is expected to solve this issue. For example, in loop quantum cosmology, many studies have successfully avoided these singularities, specifically referencing \cite{Ashtekar:2006rx,Ashtekar:2006wn,Ashtekar:2006uz,Ashtekar:2003hd,Papanikolaou:2023crz,Date:2004fj,Vereshchagin:2004uc,Singh:2003au,Wilson-Ewing:2016yan}.
In the context of quantum cosmology, many black hole solutions have been developed \cite{Bodendorfer:2019nvy,Ashtekar:2018lag,Brahma:2020eos,Bodendorfer:2019jay}. Recently, in the study of spherically symmetric matter collapse, scholars Lewandowski, Ma, and Yang successfully derived a quantum corrected-black hole(QCBH) model in LQG theory \cite{Lewandowski:2022zce}. This model is a modification of the Schwarzschild black hole(SBH) and also resolves the singularity problem of matter collapse. This is because, when the density of the collapsing matter reaches the Planck scale, the collapse process does not continue but instead halts and enters a bounce expansion phase \cite{Lewandowski:2022zce}. Some scholars have studied the properties of QCBH, such as the shadow, photon ring, and quasinormal modes \cite{Gong:2023ghh,Ye:2023qks,Yang:2022btw,Afrin:2022ztr}. Additionally, other aspects of QCBH have been explored in the literature \cite{Shao:2023qlt,Zhang:2023yps,Giesel:2022rxi}. These studies have investigated different characteristics of QCBH, and further exploration of the lensing effects of QCBH would be an interesting research direction.

Gravitational lensing, used as an astronomical observation tool, occurs when massive objects such as galaxies or black holes distort the surrounding spacetime, causing the path of light to bend. This phenomenon was first confirmed through the deflection of sunlight observed by Eddington and others during the 1919 solar eclipse \cite{Dyson:1920cwa}. This phenomenon was later further theorized and proposed for astronomical observation applications by researchers such as Refsdal and Liebes \cite{Liebes:1964zz,10.1093/mnras/128.4.295}.
Subsequently, the application of gravitational lensing has extended to black hole research, providing a novel means of integrating theoretical analysis with astronomical observational data. Regarding strong gravitational lensing, Virbhadra et al. were the first to derive the lens equation for a SBH in the strong-field limit through numerical analysis \cite{Virbhadra:1999nm}. In 2001, Bozza et al. derived the theoretical lensing formula for SBH in the strong-field limit and subsequently extended it to general static spherically symmetric spacetimes the following year \cite{Bozza:2001xd,Bozza:2002zj}. Finally, this method was refined and extended to general asymptotically flat spacetimes \cite{Tsukamoto:2016jzh} and axially symmetric spacetimes \cite{Duan:2023gvm,Kumar:2023jgh,Ghosh:2022mka,Hsieh:2021rru,Islam:2021dyk,Hsieh:2021scb,Chen:2011ef,Ji:2013xua}. 
Based on the method proposed by Bozza, this method has been applied to Reissner-Nordström black hole and braneworld black hole \cite{Eiroa:2003jf,Whisker:2004gq,Eiroa:2005vd,Li:2015vqa}.
Additionally, corresponding studies have been conducted on black holes in other spacetime backgrounds \cite{Islam:2020xmy,Bozza:2003cp,Kumar:2019pjp,Kumar:2021cyl,QiQi:2023nex,Panpanich:2019mll,Babar:2021nst,Kuang:2022xjp,Soares:2023uup,Soares:2023err}.

In fact, strong gravitational lensing, as an essential tool for studying black holes and cosmology, has been extensively researched and applied in cosmology, astronomy, and physics \cite{Li:2015vqa}. Particularly, the breakthroughs made by the EHT have opened the door to directly observing strong gravitational fields \cite{EventHorizonTelescope:2019uob,EventHorizonTelescope:2022wkp}, providing an unprecedented perspective for studying astronomical objects such as black holes. This achievement has made the strong gravitational lensing effect a research hotspot because it allows direct observation of celestial bodies in strong gravitational environments, reveals the characteristics of black holes under different gravitational theories, and enables comparison with the predictions of GR \cite{QiQi:2023nex}. Therefore, exploring the impact of quantum correction parameter on the lensing effect in a QCBH model can further our understanding of quantum effects.
In LQG theory, researchers Lewandowski, Ma, and Yang have recently proposed an innovative QCBH model (quantum corrected-black hole) \cite{Lewandowski:2022zce}. This black hole model provides a new theoretical framework for studying gravitational lensing effects, particularly in exploring the impact of quantum correction parameter on the lensing effect.

The structure of this paper is as follows: In Section \ref{sec:level2}, we primarily review QCBH. In Section \ref{sec:level3},we use the method proposed by Bozza et al. to handle the deflection angle of gravitational lensing in the strong-field limit. We calculate the deflection angle and the corresponding deflection angle coefficients ($\bar{a}$ and $\bar{b}$) in QCBH, analyzing the impact of the quantum correction coefficient \(\alpha\) on these deflection angles and lensing coefficients. In Section \ref{sec:level4}, we focus on supermassive black holes ($M87^*$ and $Sgr A^*$) to analyze lensing observations, Einstein rings, and time delays. Additionally, we constrain the quantum correction parameter using the EHT observations of the shadows of supermassive black holes $M87^*$ and $Sgr A^*$. The final section of the paper provides a summary and prospect. Throughout the entire paper, we use natural units, i.e., $c=\hbar=G=1$.

\section{\label{sec:level2}Quantum-corrected black hole}
Regarding the QCBH derived by Lewandowski, Ma, and Yang \cite{Lewandowski:2022zce}, its metric is as follows:
\begin{equation}
ds^2=-f(r)dt^2+\frac{1}{f(r)}dr^2+r^2(d\theta^2+sin^2\theta d \phi^2),
\label{1}
\end{equation}
where
\begin{equation}
f(r)=1-\frac{2M}{r}+\frac{\alpha M^2}{r^4}.
\label{2}
\end{equation}
Here, $\alpha=16\sqrt{3}\gamma^3\ell^2_p $ is the quantum correction parameter, $\ell^2_p=1$, and $M$represents the mass of the QCBH.
It is worth noting that when the quantum correction parameter $\alpha$ vanishes ($\alpha=0$), the QCBH degenerates into the SBH ($f(r)=1-\frac{2M}{r}$).
Analyzing the metric (\ref{2}), it is easy to see that $\displaystyle\lim_{x \to \infty} f(r)\to 1$, indicating that the spacetime is asymptotically flat. For convenience in discussing gravitational lensing, the line element (\ref{1}) can be rewritten using dimensionless parameter transformations as:
\begin{equation}
d\tilde{s}^2 = -A(x)dT^2 + \left( \frac{1}{B(x)} \right) dx^2 + C(x) \left( d\theta^2 + \sin^2 \theta d\phi^2 \right),
\label{3}
\end{equation}
where
\begin{equation}
A(x) = \frac{1}{B(x)} = 1 - \frac{1}{x} + \frac{\tilde{\alpha}}{16x^4},
\label{4}
\end{equation}
and
\begin{equation}
C(x) = x^2.
\label{5}
\end{equation}
The dimensionless parameters are defined as:
\begin{equation}
r = 2Mx, \quad t = 2MT, \quad \alpha = \tilde{\alpha}M^2.
\label{6}
\end{equation}

The analytical expression for the event horizon of the black hole can be obtained from the condition $g^{rr} = 0$, that is,
\begin{equation}
\bm{A(x)=1-\frac{1}{x}+\frac{\tilde{\alpha}}{16x^4} =0.}
\label{7}
\end{equation}
Clearly, when the quantum correction parameter takes different values, the number of roots can be zero, one, or two, corresponding physically to the non-existence of a black hole event horizon, the existence of one event horizon, and the existence of two event horizons, respectively. In this article, the primary discussion is on the case where an event horizon exists. 
Therefore, the quantum correction parameter is constrained within the range $ \frac{27}{16} \geq \tilde{\alpha} \geq 0$, which aligns with the discussion of black hole existence (at least one event horizon).

\begin{figure}[]
\includegraphics[width=0.5 \textwidth]{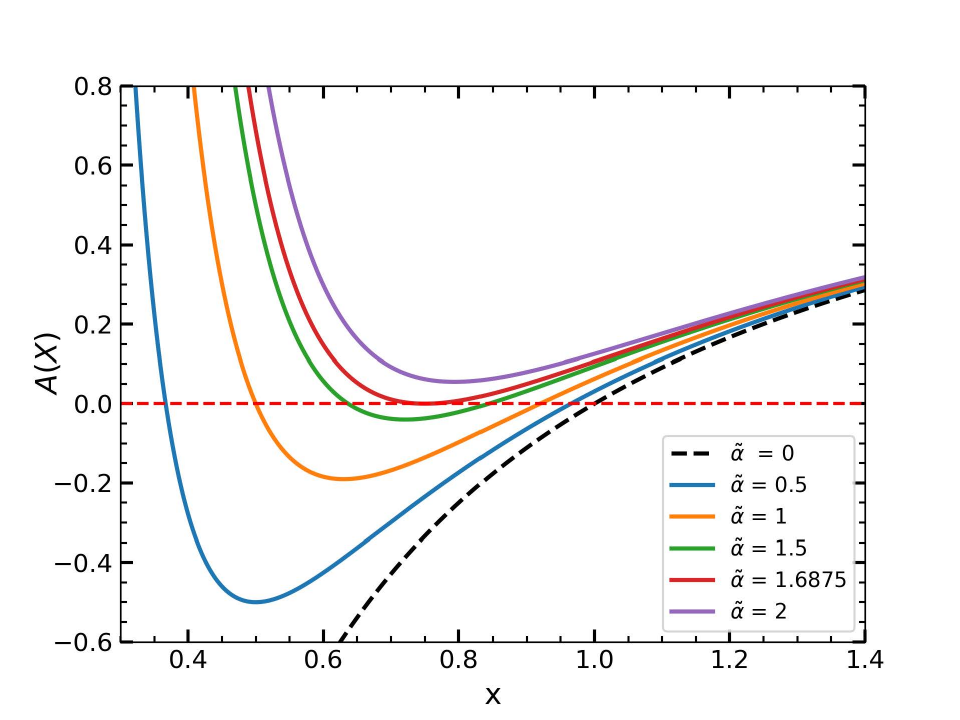}
\caption{
Different quantum correction parameters affect the existence of the event horizon of QCBH, with the black dashed line representing the event horizon of a SBH.}
\label{a}
\end{figure}

As shown in Figure \ref{a}, when the quantum correction parameter $\tilde{\alpha}$ gradually increases, the number of event horizons also changes significantly. This indicates that the quantum correction parameter has a significant impact on the properties of QCBH. 
\textbf{When the quantum correction parameter increases to a value exceeding the critical threshold, $A(x)$ has no real roots, which physically means that there is no black hole in this spacetime.}
By analyzing Figure \ref{a}, it is evident that when the quantum correction parameter is absent ($\tilde{\alpha} = 0$), metric (\ref{4}) represents a standard SBH. It is clear that the event horizon radius of the QCBH is smaller than that of the SBH (the black dashed line in the figure represents the SBH).

\section{\label{sec:level3}Strong gravitational lensing effect}
In this section, we use the strong field limit method by V. Bozza et al. to calculate the deflection angle near the unstable photon sphere \cite{Bozza:2002zj}. This method is an extension of the work presented in \cite{Bozza:2001xd} and provides a general approach for extending the strong field limit to arbitrary static spherically symmetric spacetimes. 
In the second section, we analyzed QCBH and found that it is asymptotically flat, thus this method is applicable in the spacetime of QCBH.
In QCBH, considering it is static and spherically symmetric, for the convenience of analysis, photons can be restricted to the equatorial plane, i.e.,$\theta = \frac{\pi}{2}$.
In this case, metric (\ref{3}) becomes

\begin{equation}
d{\widetilde{s}}^2=-A(x)dT^2+\frac{1}{B(x)}dx^2+C(x)d\phi^2.
\label{8}
\end{equation}

In stable and spherically symmetric spacetime structures, the four-momentum of photons along directions that preserve time and spatial symmetries (Killing vector fields) is conserved. Therefore, the energy $E$ and angular momentum $L$ of a photon are related to the Killing vector fields $\xi_t^\mu$ and $\xi_\phi^\mu$,  which are associated with time translation symmetry and axial rotational symmetry, respectively. That is, the energy of the photon is defined as $E=-p_\mu\xi_t^\mu$, and the angular momentum of the photon is defined as $L={p_\mu\xi}_\phi^\mu$, where $p_\mu$ are the components of the photon's four-momentum.Therefore, we obtain
\begin{equation}
\frac{d\phi}{d\lambda}=\frac{L}{C(x)},
\label{9}
\end{equation}

\begin{equation}
\frac{dt}{d\lambda}=-\frac{E}{A(x)}.
\label{10}
\end{equation}
Here, $\lambda$ is an affine parameter. We are primarily concerned with the deflection of light rays as they approach the surface of the photon sphere. During this process, the geodesic motion of the light rays satisfies the null geodesic condition, i.e., $d{\widetilde{s}}^2= 0$. Combining the metric (\ref{8}) with equations (\ref{9}) and (\ref{10}), we get
\begin{equation}
-A(x)\frac{E^2}{A(x)^2}+\frac{1}{A(x)}{(\frac{dx}{d\lambda})}^2+C(x)\frac{L^2}{{C(x)}^2}=0.
\label{11}
\end{equation}
Rearranging the above equation, we obtain
\begin{equation}
{(\frac{dx}{d\lambda})}^2=E^2-\frac{L^2 A(x)}{C(x)}.
\label{12}
\end{equation}
The path of a photon moving around a black hole can be described using an effective potential \cite{Pugliese:2010ps,Gan:2021pwu,Guo:2022muy,Mustafa:2022xod}. The radial effective potential can be given by the following expression
\begin{equation}
V_{eff}(x)=\frac{L^2 A(x)}{C(x)}=\frac{L^2}{x^2}(1-\frac{1}{x}+\frac{\widetilde{\alpha}}{{16x}^4}).
\label{13}
\end{equation}

\begin{figure}[]
\includegraphics[width=0.5 \textwidth]{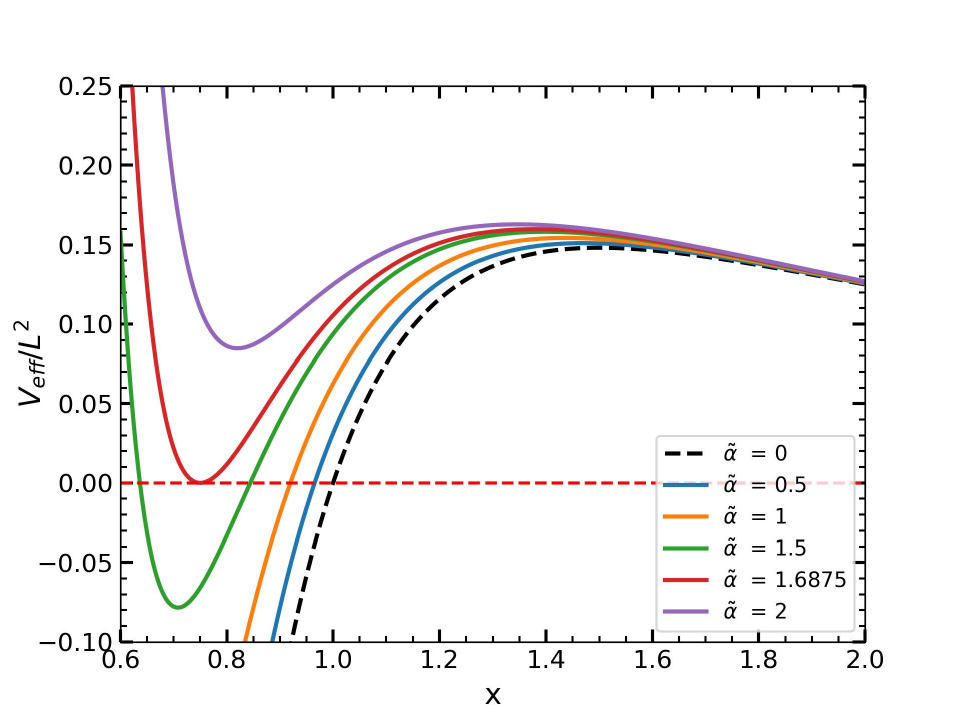}
\caption{
The trend of the effective potential graph. From the graph, it can be intuitively seen that there exist two photon spheres. The one with the smaller radial distance represents the stable photon sphere, while the one with the larger radial distance represents the unstable photon sphere.}
\label{b}
\end{figure}

According to the radial effective potential, for light rays coming from infinity and incident on the black hole, when the light rays reach the vicinity of the black hole, due to the presence of the effective potential, the light rays can be deflected at a specific radius $x_0$ (this distance is the closest approach of the photon to the black hole). At this position, the photon will not fall into the black hole but will escape from the black hole and symmetrically return to infinity to be observed by an observer. These orbital radii can be derived from the expression of the effective potential, mathematically described as
\begin{equation}
\frac{dV_{eff}(x)}{dx}=0 \;\; \text{photon sphere},
\label{14}
\end{equation}

\begin{equation}
\frac{dV_{eff}(x)}{dx}=0  \;  \text{and} \; \frac{d^2V_{eff}(x)}{dx^2}<0 \; \text{unstable photon sphere},
\label{15}
\end{equation}

\begin{equation}
\frac{dV_{eff}(x)}{dx}=0 \; \text{and} \; \frac{d^2V_{eff}(x)}{dx^2}>0\; \text{stable photon sphere}.
\label{16}
\end{equation}
The solutions of the above equation correspond to the radii of stable photon spheres or unstable photon spheres. Obviously, without loss of generality, we are more interested in the unstable photon sphere and, on this basis, study the behavior of light deflection in the strong field limit. Therefore, when considering only the orbital radius of the unstable photon sphere, according to equation (\ref{15}), we obtain
\begin{equation}
\frac{{A(x)}^{'}}{A(x)}=\frac{{C(x)}^{'}}{C(x)}.
\label{17}
\end{equation}
Substituting equations (\ref{4}) and (\ref{5}) into the above equation, we obtain
\begin{equation}
16x^4-24x^3+3\widetilde{\alpha}=0.
\label{18}
\end{equation}
The roots of equation (\ref{18}) represent the radius of the photon sphere. Obviously, since equation (\ref{18}) is a quartic equation, there will be two roots regardless of the value of the quantum correction parameter. However, carefully analyzing the other condition of equation (\ref{15}), it is evident that the radius of the unstable photon sphere corresponds to the largest root. This is because at the largest root, the motion of the photon satisfies condition (\ref{15}). This can be visually observed from the effective potential (as shown in Figure \ref{b}). According to the trend of the effective potential, the unstable photon sphere is located at a larger radial distance. Therefore, for equation (\ref{18}), taking the largest root as the unstable orbital radius of the photon $x = x_m$ is most appropriate. In the following discussion, $x_m$ will represent only the orbital radius of the unstable photon sphere.

When a light ray travels from infinity to the vicinity of a black hole, it carries a certain impact parameter $b$. During this process, the light ray approaches the black hole at a minimum distance $x_0$ and is then symmetrically deflected back to infinity. The relationship between the impact parameter $b$ and the minimum distance $x_0$ the light ray reaches near the black hole can be obtained from $V_{eff}(x) = E^2$ (where the radial motion is zero). Combining equation (\ref{13}), the impact parameter can be expressed as:
\begin{equation}
b = \frac{L}{E} = \sqrt{\frac{C(x_0)}{A(x_0)}} = \frac{4x^3}{\sqrt{\tilde{\alpha} + 16x^4 - 16x^3}}.
\label{19}
\end{equation}

For the radius $x_m$ of the unstable photon sphere, choosing $x_0 = x_m$, the corresponding impact parameter is $b_m$. In the strong field limit, the deflection angle of light can be given by the definition in the literature \cite{Virbhadra:1999nm},that is,
\begin{equation}
\alpha_D(x_0)=I(x_0)-\pi,
\label{20}
\end{equation}
where
\begin{equation}
I(x_0)=2\int_{x_0}^{\infty}\frac{d\phi}{dx}=2\int_{x_0}^{\infty}\frac{1}{\sqrt{A(x)C(x)}\sqrt{\frac{A(x_0)C(x)}{C(x_0)A(x)}-1}}dx.
\label{21}
\end{equation}
The detailed derivation of the above expression can be found in the literature \cite{Virbhadra:1998dy}.

To calculate the above expression, we use the approximation method from the literature \cite{Bozza:2002zj}, expanding the deflection angle near the photon sphere. For this purpose, a new variable is redefined \cite{Tsukamoto:2016jzh,Jha:2023qqz,Fu:2021fxn,Islam:2021dyk}:
\begin{equation}
z = 1 - \frac{x_0}{x} .
\label{22}
\end{equation}
Using this variable, the integral (\ref{21}) can be rewritten as
\begin{equation}
I(x_0) = \int_0^1 R(z, x_0) f(z, x_0)dz,
\label{23}
\end{equation}
where $R(z, x_0)$ can be expressed as
\begin{equation}
R(z, x_0) = \frac{2x_0 \sqrt{C(x_0)}}{C(x) (1 - z)^2},
\label{24}
\end{equation}
and $f(z, x_0)$ can be expressed as
\begin{equation}
f(z, x_0) = \frac{1}{\sqrt{A(x_0) - \frac{A(x) C(x_0)}{C(x)}}}.
\label{25}
\end{equation}
It is easy to see that the integrals for all values of the function \(R(z, x_0)\) are regular, but the function \(f(z, x_0)\) diverges at \(z = 0\). Therefore, to avoid the divergence at \(z = 0\), the function \(f(z, x_0)\) can be expanded in a series at \(z = 0\), retaining the first and second-order approximations as
\begin{equation}
f(z, x_0) \approx f_0(z, x_0) = \frac{1}{\sqrt{\gamma_1(x_0)z + \gamma_2(x_0)z^2}},
\label{26}
\end{equation}
where the parameters $\gamma$ can be read as
\begin{equation}
\gamma_1\left(x_0\right)=\frac{x_0}{C\left(x_0\right)}\left[C^{'}\left(x_0\right)A\left(x_0\right)-A^{'}\left(x_0\right)C\left(x_0\right)\right],
\label{27}
\end{equation}
and
\begin{align}
\gamma_2(x_0)=&\frac{1}{2}\left[\frac{2x^2_0 C^{'}(x_0)\left(A^{'}(x_0)C(x_0)-C^{'}(x_0)A(x_0)\right)}{C(x_))^2} \right.\notag\\
&\left.+\frac{x_0}{C\left(x_0\right)}(C^{''}\left(x_0\right)A\left(x_0\right)-A^{''}\left(x_0\right)C\left(x_0\right))\right].
\label{28}
\end{align}
According to the method used in the reference \cite{Bozza:2002zj}, the integral can be divided into two parts: one part is divergent, and the other part is regular. Therefore, it can be written as
\begin{equation}
I(x_0)=I_D(x_0)+I_R(x_0),
\label{29}
\end{equation}
the divergent part $I_D(x_0)$ is expressed as
\begin{equation}
I_D(x_0)=\int_{0}^{1}{R(0,x_m)f_0((z,x_0)}dz,
\label{30}
\end{equation}
the regular part $I_R(x_0)$ is expressed as
\begin{equation}
I_R(x_0)=\int_{0}^{1}\left(R(z,x_0)f((z,x_0)-R(0,x_m)f_0((z,x_0)\right)dz.
\label{31}
\end{equation}
$I_D(x_0)$ represents the regular part after subtracting the divergent part of the integral. Therefore, solve the above two integrals \eqref{30} and \eqref{31}. Near $x_m$ the deflection angle of light in the strong field limit can be expressed as \cite{Bozza:2003cp,Bozza:2002zj}
\begin{equation}
\alpha_D(b)=-\bar{a}log(\frac{b}{b_m}-1)+\bar{b}+O(b-b_m).
\label{32}
\end{equation}
The corresponding coefficients can be written as
\begin{equation}
\bar{a}=\frac{R(0,x_m)}{2\sqrt{\gamma_2\left(x_m\right)}},
\label{33}
\end{equation}
\begin{equation}
\bar{b}=-\pi+I_R(x_m)+\bar{a}log(\frac{2\gamma_2\left(x_m\right)}{G(x_m)}).
\label{34}
\end{equation}

\begin{figure*}[ht]
\includegraphics[width=1 \textwidth]{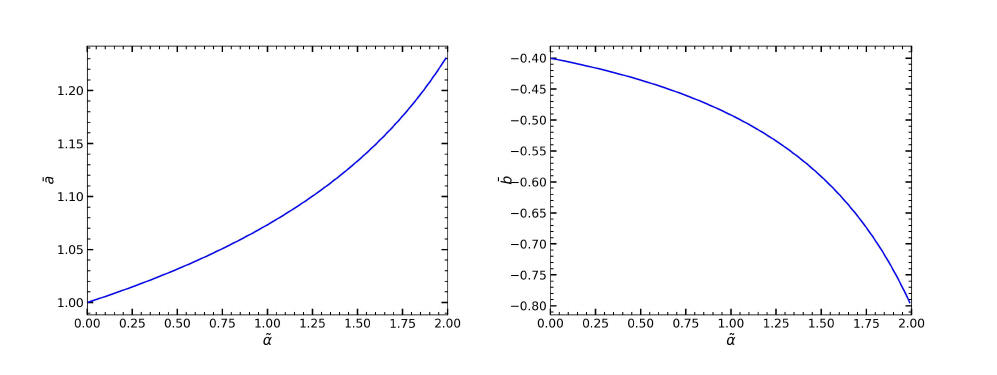}
\caption{
Left image: The variation of the deflection coefficient \(\bar{a}\) with the quantum correction parameter \(\tilde{\alpha}\) in a strong field.  
Right image: The variation of the deflection angle coefficient \(\bar{b}\) with the quantum correction parameter \(\tilde{\alpha}\).}
\label{c}
\end{figure*}

\begin{figure*}[ht]
\includegraphics[width=1 \textwidth]{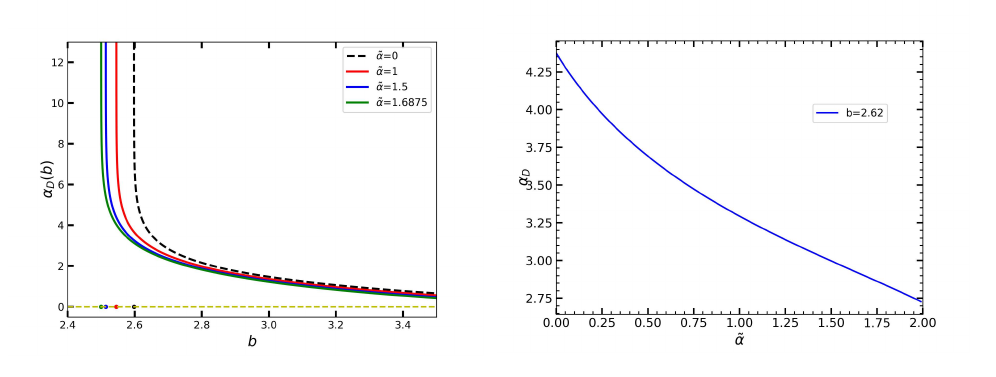}
\caption{
Left image: The variation of the deflection angle with the impact parameter under different quantum correction parameters. The black dashed line represents the deflection angle for the SBH.  
Right image: The variation of the deflection angle with the quantum correction parameter at the impact parameter \(b=2.62\).}
\label{d}
\end{figure*}

Here, we numerically solve to characterize the relationship between the strong gravitational lensing coefficients and the quantum correction parameter. As shown in Figure \ref{c}, it is evident that the deflection coefficient \(\bar{a}\) gradually increases with the increase of the quantum correction parameter, while the deflection coefficient \(\bar{b}\) gradually decreases with the decrease of the quantum correction parameter \(\tilde{\alpha}\). It is worth mentioning that when the quantum correction parameter vanishes (\(\tilde{\alpha} = 0\)), the QCBH becomes a SBH. Our results match the values for the SBH \cite{Bozza:2002zj}, i.e., \(\bar{a} = 1\), \(\bar{b} = -0.40023\) (see Table \ref{table1}). In Figure \ref{d}, under different quantum correction parameters, the deflection angle diverges at certain values (\(b = b_m\)). As the quantum correction parameter increases, the corresponding divergent impact parameter gradually decreases, and the deflection angle also significantly decreases (see the left panel of Figure \ref{d}). Naturally, for the same impact parameter, the deflection angle of the SBH is significantly greater than that of the QCBH and decreases with the increase of the quantum correction parameter (see the right panel of Figure \ref{d}).

\section{\label{sec:level4}Gravitational Lensing Effects of Supermassive Black Holes and Constraints from the EHT}

\subsection{\label{sec:level4.1}Characteristic observables in strong lensing effect}

In Section \ref{sec:level3}, the deflection angle for strong gravitational lensing was calculated. Therefore, the position of the image can be easily determined using the lens equation here. According to the definition of the lens equation in the literature \cite{Virbhadra:1999nm,Bozza:2008ev,Bozza:2001xd}, the lens equation can be easily obtained as
\begin{equation}
\beta=\theta-\frac{D_{LS}}{D_{OS}}\Delta\alpha_n.
\label{35}
\end{equation}
Here, $D_{LS}$ is the distance between the lens and the light source, and $D_{OS}$  is the distance between the observer and the light source $(D_{OS}=D_{OL}+D_{LS})$, $\beta$ and $\theta$  represent the angular positions of the source and image relative to the optical axis, and $\Delta\alpha_n=\alpha(\theta)-2n\pi$ denotes the deflection of light after orbiting the black hole $n$ times. To approximate the deflection $\Delta\alpha_n$, we need to find the angle $\theta_n^0$, which is obtained by solving $\alpha(\theta)=2n\pi$. Our adopted solution is given by the following equations
\begin{equation}
\theta_n^0=\frac{b_m(1+e_n)}{D_{OL}},
\label{36}
\end{equation}
where
\begin{equation}
e_n=exp(\frac{\bar{b}-2n\pi}{\bar{a}}).
\label{37}
\end{equation}

Next, by combining the deflection angle formula \eqref{36} in the strong field limit and the gravitational lens equation \eqref{35}, while neglecting higher-order terms, we can approximate the position of the nth image \cite{Bozza:2002zj}
\begin{equation}
\theta_n=\theta_n^0+\frac{b_me_n(\beta-\theta_n^0)D_{OS}}{\bar{a}D_{LS}D_{OL}}.
\label{38}
\end{equation}
From the above equation, it can be seen that when \(\beta - \theta_n^0 = 0\), the image position coincides with the source position. Clearly, at this moment, the image position is \(\theta_n = \theta_n^0\), which means that the position of the \(n\)-th image has not been corrected (indicating that the source and the image are on the same side). To obtain the position of the image on the opposite side of the source, an extension is made by replacing \(\beta\) with \(-\beta\). This way, the position of the \(n\)-th image on the opposite side of the source is obtained. It is worth noting that when the light, the lens (black hole), and the observer are aligned, i.e., \(\beta = 0\), solving equation (\ref{38}) can yield
\begin{equation}
\theta_n^E=(1-\frac{b_me_nD_{OS}}{\bar{a}D_{LS}D_{OL}})\ \theta_n^0.
\label{39}
\end{equation}
This is known as the Einstein ring \cite{Einstein:1936llh}. For the relativistic image with \(n=1\) (\(\theta_1^E\)) and when the black hole is located between the observer and the source (with \(D_{OS} = D_{LS} = 2D_{OL}\)). Considering the case where \(D_{OL}\) is much larger than the impact parameter \(b_m\) (\(D_{OL} \gg b_m\)), and combining with equation (\ref{36}), we obtain
\begin{equation}
\theta_1^E = \frac{(1 + e_1)b_m}{D_{OL}},
\label{40}
\end{equation}
here
\begin{equation}
e_1 = \exp\left(\frac{\bar{b} - 2\pi}{\bar{a}}\right).
\label{41}
\end{equation}

Apart from the position of the source image, its magnification is also an important piece of information. Therefore, the magnification of the \(n\)-th image can be defined as \cite{Bozza:2002zj,Virbhadra:1998dy,Virbhadra:2007kw}
\begin{equation}
\mu_n=\left.\left(\frac{\beta}{\theta}\frac{d\beta}{d\theta}\right)^{-1}\right|_{\theta_n^0}=\frac{{b_m}^2(1+e_n)D_{OS}}{\bar{a}\beta D_{LS}{D_{OL}}^2}e_n.
\label{42}
\end{equation}
From the above equation, it is clear that the magnification factor decreases with the increase in image layer number \( n \) and decays exponentially. When the parameter \( \beta \) approaches zero, the magnification factor reaches its maximum, making the relativistic images the brightest and thus the easiest to observe.
To simplify observations, typically only two layers of images are analyzed: the outermost image \(\theta_1\) and all inner images considered as a whole \(\theta_{\infty}\). Through this simplification, some interesting observational results can be obtained.

The position of all inner images considered as a whole, denoted as $\theta_\infty$, is
\begin{equation}
\theta_\infty=\frac{b_m}{D_{OL}},
\label{43}
\end{equation}
the separation $s$ between the first image and the other images is
\begin{equation}
s=\theta_1-\theta_\infty=\theta_\infty exp(\frac{\bar{b}-2\pi}{\bar{a}}),
\label{44}
\end{equation}
the flux ratio of the brightness between the first image and the other images is
\begin{equation}
r=\frac{\mu_1}{\sum_{n=2}^{\infty}\mu_n}=exp(\frac{2\pi}{\bar{a}}),\, r_{mag}\approx2.5log(r)=\frac{5\pi}{\bar{a}ln(10)}.
\label{45}
\end{equation}
From equation \eqref{45}, it can be seen that, in this case, the flux ratio is independent of the distance between the lens and the observer.

Using the above equations (\ref{43}), (\ref{44}), and (\ref{45}), as long as the lensing coefficients \(\bar{a}\) and \(\bar{b}\), as well as the critical impact parameter \(b_m\), can be determined, the observational values of the QCBH under strong gravitational lensing can be theoretically calculated. Conversely, astronomical observations can enhance our understanding of QCBH properties.

Of course, in the study of strong gravitational lensing effects, time delay is also an important observable. Time delay mainly reflects the time difference experienced by photons traveling along different paths around a black hole (considering that light rays near the photon sphere may orbit the black hole several times). Since the path lengths and travel times of these light rays are different, this results in time delays between the formed relativistic images. This phenomenon of time delay can be obtained through astronomical observations and has been widely studied in the field of astrophysics. For example, it can be used to estimate the Hubble constant parameter \cite{Grillo:2024rhi,Birrer:2022chj,Qi:2022sxm,Treu:2022aqp,Grillo:2018ume}.

For relativistic images located on the same side of the lens (black hole), the time delay can be obtained from the literature \cite{Bozza:2003cp}. In this case, the time delay between relativistic images can be written as
\begin{equation}
\Delta T_{n,l} = 2\pi(n-l) \left(\frac{\tilde{a}}{\bar{a}}\right) + 2\sqrt{\frac{A_m b_m}{B_m}} \sqrt{b_m} \left(e^{-\frac{\bar{b} - 2n\pi}{2\bar{a}}} - e^{-\frac{\bar{b} - 2l\pi}{2\bar{a}}}\right).
\label{46}
\end{equation}
Here, the first term reflects the time delay caused by photons orbiting the black hole different numbers of times. The second term is mainly a correction term for the time delay (due to the time dilation effect of light in the gravitational field). In exploring the impact of time delay, it is evident that the first term dominates. Since the QCBH discussed in this paper is static and spherically symmetric, the above equation can be rewritten as
\begin{equation}
\Delta T_{n,l} \approx 2\pi(n-l) \frac{\tilde{a}}{\bar{a}} = 2\pi(n-l) u_m = 2\pi(n-l) \theta_\infty D_{OL}.
\label{47}
\end{equation}
Based on equation (\ref{47}), the time delay between two relativistic images can be calculated precisely. If future observational technology can accurately distinguish these images, precise time delay data can be obtained in astrophysical observations. This is crucial for a deeper understanding of black holes and their quantum effects. Additionally, by further analyzing multiple relativistic images, the properties of the QCBH can be better understood. With the development of high-resolution astronomical observation technology, achieving such precision is only a matter of time.

\subsection{\label{sec:level4.2}Lensing effects of the supermassive black holes M87* and SgrA*}
To evaluate several interesting observational values calculated in the previous section, in this section, the QCBH will be considered as the supermassive black holes $M87^*$ and $Sgr A^*$. This will be used to study these observable values, and the simulated data will be compared with those of the SBH (when the quantum correction parameter vanishes, the QCBH degenerates into a SBH).

According to the latest astronomical observational data, we know that the mass of M87* is \((6.5 \pm 0.7) \times 10^9 M_\odot\), and its distance from Earth is \((16.8 \pm 0.8)\) Mpc \cite{EventHorizonTelescope:2019ggy}. The mass of SgrA* is \(4^{+1.1}_{-0.6} \times 10^6 M_\odot\), and its distance from Earth is \(8.15 \pm 0.15\) kpc \cite{Chen:2019tdb,EventHorizonTelescope:2022xqj}. Through these observational data, it is easy to see how the quantum correction parameter \(\tilde{\alpha}\) affects the SBH, and further explore the properties of the quantum correction parameter.

\begin{figure*}[ht]
\includegraphics[width=1 \textwidth]{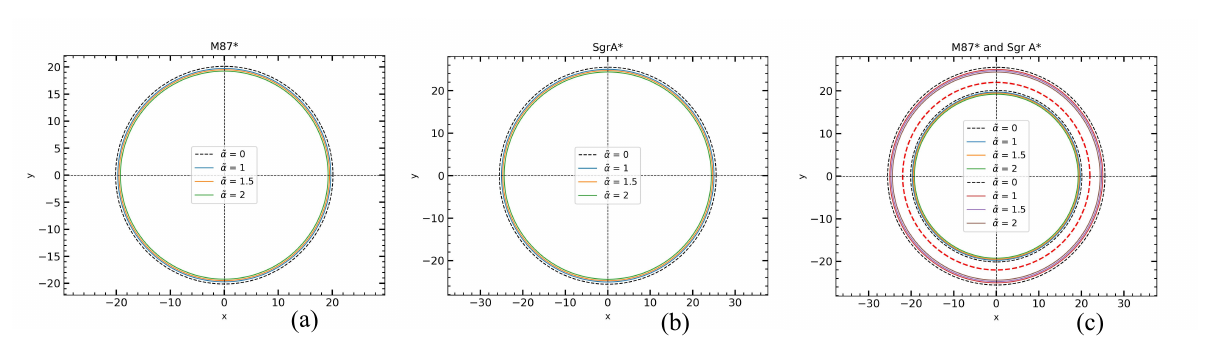}
\caption{
For \(n=1\), considering the QCBH as the supermassive $M87^*$ and $Sgr A^*$ black holes' Einstein rings. Figure a represents the Einstein ring in the context of the $M87^*$ black hole, Figure b represents the Einstein ring in the context of the $Sgr A^*$ black hole, and Figure c shows the Einstein rings for both $M87^*$ and $Sgr A^*$ black holes. The black dashed line in the figures represents the case where the QCBH degenerates into a SBH. The Einstein ring outside the red dashed line corresponds to the $Sgr A^*$ black hole, while the Einstein ring inside the red dashed line corresponds to the $M87^*$ black hole.}
\label{e}
\end{figure*}

Considering the QCBH as representative of the supermassive black holes $M87^*$ and $Sgr A^*$, and studying their Einstein rings accordingly.
As shown in Figure \ref{e} and Table \ref{table1}, the quantum correction parameter does not significantly affect the size of the Einstein ring. The black dashed line in the figure represents the situation where the quantum correction parameter is absent, at which point the QCBH degenerates into SBH (\(\tilde{\alpha} = 0\)).
For both the $M87^*$ black hole and the $Sgr A^*$ black hole, the effect of the quantum correction parameter on the Einstein ring is to reduce its size (see Figures \ref{e}a and \ref{e}b). In Figure \ref{e}c, it is evident that the size of the Einstein ring in the context of the $Sgr A^*$ black hole is significantly larger than that in the context of the $M87^*$ black hole. This phenomenon persists even when considering the quantum correction parameter and does not disappear.
This can be well explained physically. Firstly, the effect of the quantum correction parameter is not very sensitive, so its presence does not cause significant changes in the size of the Einstein ring. Secondly, astronomical observation data show that the $SgrA^*$ black hole is closer to Earth, making its Einstein ring appear larger from our perspective, while the Einstein ring of the $M87^*$ black hole appears smaller due to its greater distance.

Furthermore, in the studies using $M87^*$ and $SgrA^*$ black holes as backgrounds, for the $M87^*$ black hole, the quantum correction parameter causes the deviation in the Einstein ring between the QCBH and the SBH to be below $0.7256 \mu as$. For the $SgrA^*$ black hole, this deviation is below $0.9205 \mu as$ (see Table \ref{table1}). Clearly, with future upgrades in observational equipment, these differences will be detectable. This is crucial for further understanding the properties of the quantum correction parameter.

\begin{table*}[]
\centering
\begin{tabular}{p{1.5cm}p{1.5cm}p{1.5cm}p{1.5cm}p{2.5cm}p{1.5cm}p{1.5cm}p{1.5cm}}
\hline\hline
& \multicolumn{2}{l}{Lensing Coefficients} & \multicolumn{1}{c}{$M87^*$} & \multicolumn{1}{l}{SgrA*} & & \multicolumn{1}{l}{$M87^*$} & \multicolumn{1}{l}{SgrA*}\\
\cline{2-3}
\rule{0pt}{10pt}
$\tilde{\alpha}$& $\bar{a}$& $\bar{b}$& $\theta_1^E(\mu as)$ & $\theta_1^E(\mu as)$ & $\delta \tilde{\alpha}$ & $\delta\theta_1^E(\mu as)$ & $\delta \theta_1^E(\mu as)$ \\
\hline
\rule{0pt}{11pt} 
0&	1.0000&	-0.4002&	20.1293&	25.5345&	0&	&	\\
\rule{0pt}{11pt}
0.3 & 1.0179 & -0.4195 & 20.0172 & 25.3922 & 0.3 & -0.1121 & -0.1423  \\
\rule{0pt}{11pt}
 0.6 & 1.0389 & -0.4447 & 19.7741 & 25.2424 & 0.6 & -0.2302 & -0.2921  \\ 
 \rule{0pt}{11pt}
0.9 & 1.0639 & -0.4782 & 19.7741 & 25.0839 & 0.9 & -0.3552 & -0.4506  \\ 
\rule{0pt}{11pt}
1.2 & 1.0945 & -0.5244 & 19.6412 & 24.9153 & 1.2 & -0.4881 & -0.6192  \\ 
\rule{0pt}{11pt}
1.5 & 1.1335 & -0.5913 & 19.4986 & 24.7344 & 1.5 & -0.6307 & -0.8001 \\ 
\rule{0pt}{11pt}
1.6875 & 1.1642 & -0.6500 & 19.4037 & 24.6140 & 1.6875 & -0.7256 & -0.9205  \\ 
\hline\hline
\end{tabular}
\caption{Values of the lensing coefficient under different quantum correction parameters and the angular distance of the Einstein ring for the black holes $M87^*$ and $Sgr A^*$ are shown. In the table, the expression $\delta(X)$ is defined as $\delta(X) = X(\text{QCBH}) - X(\text{SBH})$. This means that $\delta(X)$ represents the deviation of the Einstein ring between the QCBH and the SBH.}
\label{table1}
\end{table*}

\begin{figure*}[]
\includegraphics[width=1 \textwidth]{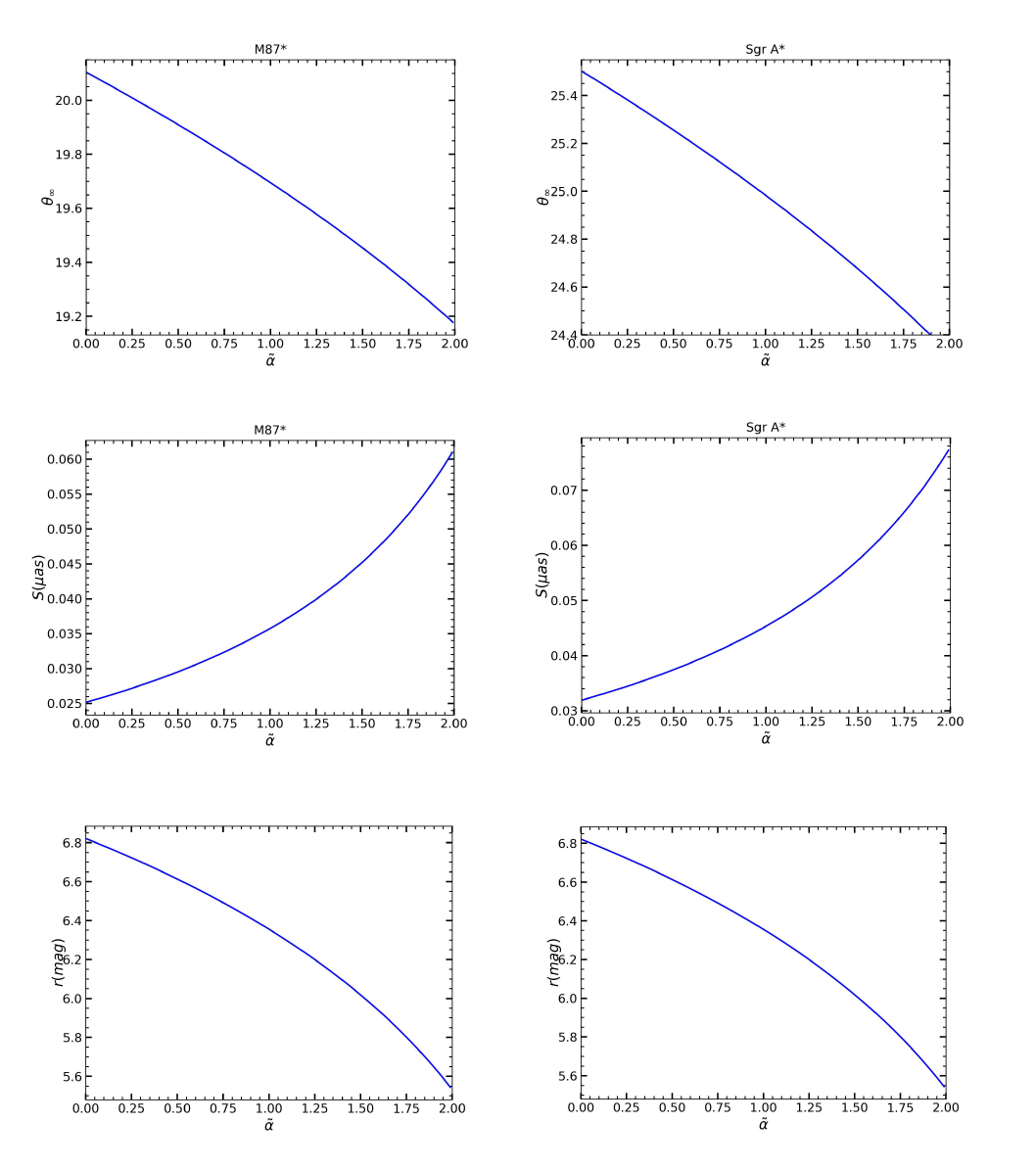}
\caption{
The variation of \( \theta_{\infty} \), \( S \), and \( r_{mag} \) with the change in the quantum correction parameter in the context of the $M87^*$ and $Sgr A^*$ black holes.
The left column represents the $M87^*$ black hole, and the right column represents the $Sgr A^*$ black hole.}
\label{f}
\end{figure*}

Using the black holes $M87^*$ and $Sgr A^*$ as the research background, simulate the observed values from the previous section (expressions (\ref{43}), (\ref{44}), and (\ref{45})) respectively.
As shown in Figure \ref{f} and Table \ref{table2}, due to the presence of the quantum correction parameter, the image position \( \theta_\infty \) decreases as the quantum correction parameter increases. The image interval \( S \) increases with the quantum correction parameter, and the brightness ratio \( r_{mag} \) between relativistic images decreases as the quantum correction parameter increases.
It is worth noting that, in the black holes $M87^*$ and $Sgr A^*$, the range of the angular position of relativistic images for the former, as the quantum correction parameter changes, is \( 20.1042 \, \mu as \geq \theta_\infty (M87*) \geq 19.3535 \, \mu as \), and for the latter, the range is \( 25.5026 \, \mu as \geq \theta_\infty (Sgr A*) \geq 24.5504 \, \mu as \). These ranges respectively match the observational ranges of the supermassive black holes $M87^*$ and $Sgr A^*$ by the EHT \cite{EventHorizonTelescope:2019dse,EventHorizonTelescope:2022wkp}.
For the deviation between the QCBH and the SBH (\( \delta(X) = X(QCBH) - X(SBH) \)), using the $M87^*$ black hole to simulate the QCBH as the background, the deviation in the angular position reaches \( |\delta(\theta_\infty)| = 0.7507 \, \mu as \) and the deviation in the image interval reaches \( |\delta(S)| = 0.025 \, \mu as \) (see Table \ref{table3}). Using the $Sgr A^*$ black hole to simulate the QCBH as the background, the deviation in the angular position reaches \( |\delta(\theta_\infty)| = 0.9522 \, \mu as \) and the deviation in the image interval reaches \( |\delta(S)| = 0.0317 \, \mu as \) (see Table \ref{table3}). These ranges all match the observational ranges of supermassive black holes by the EHT.
However, due to the resolution limitations of the EHT, which is approximately $20\mu as$ \cite{EventHorizonTelescope:2019ths}, these differences cannot be accurately resolved with existing equipment. Nevertheless, the next-generation EHT is expected to distinguish these differences. Once the two relativistic images can be resolved, it will be possible to differentiate between the SBH and the QCBH, allowing for further investigation into the properties of QCBH.

\begin{table*}[]
\centering
\begin{tabular}{p{1.5cm}p{1.5cm}p{1.5cm}p{1.5cm}p{1.5cm}p{0.5cm}p{1.5cm}p{1.5cm}p{1.5cm}p{1.5cm}}
\hline\hline
& \multicolumn{4}{c}{$M87^*$}& & \multicolumn{4}{c}{$Sgr A^*$}\\
\cline{2-5}
\cline{7-10}
\rule{0pt}{12pt}
$\tilde{\alpha}$& $\theta_{\infty}(\mu as)$& $S(\mu as)$& $r_{mag}$ &$\Delta T_{21}(h)$ & & $\theta_{\infty}(\mu as)$ & $S(\mu as)$ & $r_{mag}$ & $\Delta T_{21}(min)$ \\
\hline
\rule{0pt}{11pt} 
0&	20.1042&	0.0252	&6.8219 & 293.9829 & & 25.5026 & 0.0319	& 6.8219 & 10.8548	\\
\rule{0pt}{11pt}
0.3&	19.9895&	0.0276&	  6.7017&  292.3069& &	25.3572&	0.0350& 	6.7017&	10.7929\\
\rule{0pt}{11pt}
0.6&	19.8685&	0.0306&	6.5666&  290.5365&	&	25.2036&	0.0388&  6.5666&	10.7259	  \\ 
 \rule{0pt}{11pt}
0.9&	19.7398&	0.0343& 6.4124&	288.6555&  &	25.0404&	0.0435&	 6.4124&    10.6580	\\ 
\rule{0pt}{11pt}
1.2&	19.6022&	0.0390&	6.2329& 286.6423&  & 	24.8658&	0.0495&	6.2329&	10.5837 \\ 
\rule{0pt}{11pt}
1.5&	19.4534&	0.0452&	6.0185&	284.4669&&	24.6771&	0.0573&	6.0185&	10.5034\\ 
\rule{0pt}{11pt}
1.6875&	19.3535&	0.0502&	5.8599&	283.0063&&24.5504&	0.0636&		5.8599 &10.4495\\ 
\rule{0pt}{11pt}
1.8&	19.2905&	0.0537&	5.7527& 282.0854&&	24.4705&	0.0681&	5.7527 &	10.4155\\ 
\hline\hline
\end{tabular}
\caption{Using the supermassive black holes $M87^*$ and $Sgr A^*$ as the background, we analyze the observational values under different quantum correction parameters. Specifically, we focus on the time delay $\Delta T_{21}$ between the second relativistic image and the first relativistic image on the same side.}
\label{table2}
\end{table*}

\begin{table*}[]
\centering
\begin{tabular}{p{1.5cm}p{1.5cm}p{1.5cm}p{1.5cm}p{1.5cm}p{0.5cm}p{1.5cm}p{1.5cm}p{1.5cm}p{1.5cm}}
\hline\hline
& \multicolumn{4}{c}{$M87^*$}& & \multicolumn{4}{c}{$Sgr A^*$}\\
\cline{2-5}
\cline{7-10}
\rule{0pt}{10pt}
$\delta\tilde{\alpha}$& $\delta\theta_{\infty}(\mu as)$& $\delta S(\mu as)$& $\delta r_{mag}$ &$\delta \Delta T_{21}(h)$ & & $\delta \theta_{\infty}(\mu as)$ & $\delta S(\mu as)$ & $\delta r_{mag}$ & $\delta \Delta T_{21}(min)$ \\
\hline
\rule{0pt}{11pt} 
0&	0&	0	&0 & 0 & & 0 & 0	& 0 & 0	\\
\rule{0pt}{11pt}
0.3	&-0.1147&	0.0024&	-0.1202&	-1.6760&&	-0.1454&	0.0031&	-0.1202&	-0.0619\\
\rule{0pt}{11pt}
0.6&	-0.2357&	0.0054&	-0.2553&	-3.4464&&	-0.2990&	0.0069&	-0.2553&	-0.1289\\ 
 \rule{0pt}{11pt}
0.9&	-0.3644&	0.0091&	-0.4095&	-5.3274&&	-0.4622&	0.0116&	-0.4095&	-0.1968\\ 
\rule{0pt}{11pt}
1.2&	-0.502&	0.0138&	-0.589&	-7.3406&&	-0.6368&	0.0176&	-0.5890&	-0.2711 \\ 
\rule{0pt}{11pt}
1.5&	-0.6508&	0.0200&	-0.8034&	-9.5160&&	-0.8255&	0.0254&	-0.8034&	-0.3514\\ 
\rule{0pt}{11pt}
1.6875&	-0.7507&	0.02500&	-0.9620&	-10.9766&&	-0.9522&	0.0317&	-0.9620&	-0.4053\\ 
\rule{0pt}{11pt}
1.8&	-0.8137&	0.0285&	-1.0692&	-11.8975&&	-1.0321&	0.0362&	-1.0692&	-0.4393\\ 
\hline\hline
\end{tabular}
\caption{The deviation between the QCBH and the SBH is analyzed under different quantum correction parameters. When the quantum correction parameter \(\tilde{\alpha}=0\), the QCBH degenerates into a SBH. The deviation is uniformly expressed as \(\delta (X) = X (QCBH) - X (SBH)\).}
\label{table3}
\end{table*}

When considering QCBH as the supermassive $M87^*$ and $SgrA^*$ black holes, for the first and second relativistic images on the same side ($n=2$ and $l=1$), the time delay for the former can be as high as $293.9829$ hours, with the time delay deviation between the QCBH and the SBH reaching up to $10.9766$ hours. Such a time difference is sufficient to be observed by astronomical means (see Table \ref{table2} and Table \ref{table3}). For the latter, the time delay reaches $10.8548$ minutes, with a maximum relative deviation of $0.4393$ minutes, which is evidently too short to be observed (see Table \ref{table2} and Table \ref{table3}).
Overall, it is evident that in order to further explore the properties of QCBH, it is indeed possible to investigate the properties of QCBH in the context of the supermassive $M87^*$ black hole. This is because the time delay in its background can reach up to several hundred hours. However, this requires observational equipment capable of accurately resolving the two relativistic images. With the continuous upgrading of observational equipment, meeting such requirements is only a matter of time.

\subsection{\label{sec:4.3}Constraints on Quantum Correction Parameter from EHT Observations of $M87^*$ and $Sgr A^*$ Black Hole Shadows}
As discussed in Section \ref{sec:level3}, the radius of the photon sphere depends on the quantum correction parameters, meaning different parameters result in different photon rings. This provides an opportunity to constrain the quantum correction parameters using the EHT observations of the shadows of the supermassive black holes $M87^*$ and $SgrA^*$. In this section, we use the EHT data from the $M87^*$ and $SgrA^*$ black hole shadows to constrain the range of the quantum correction parameter.

\begin{figure}[]
\includegraphics[width=0.5\textwidth]{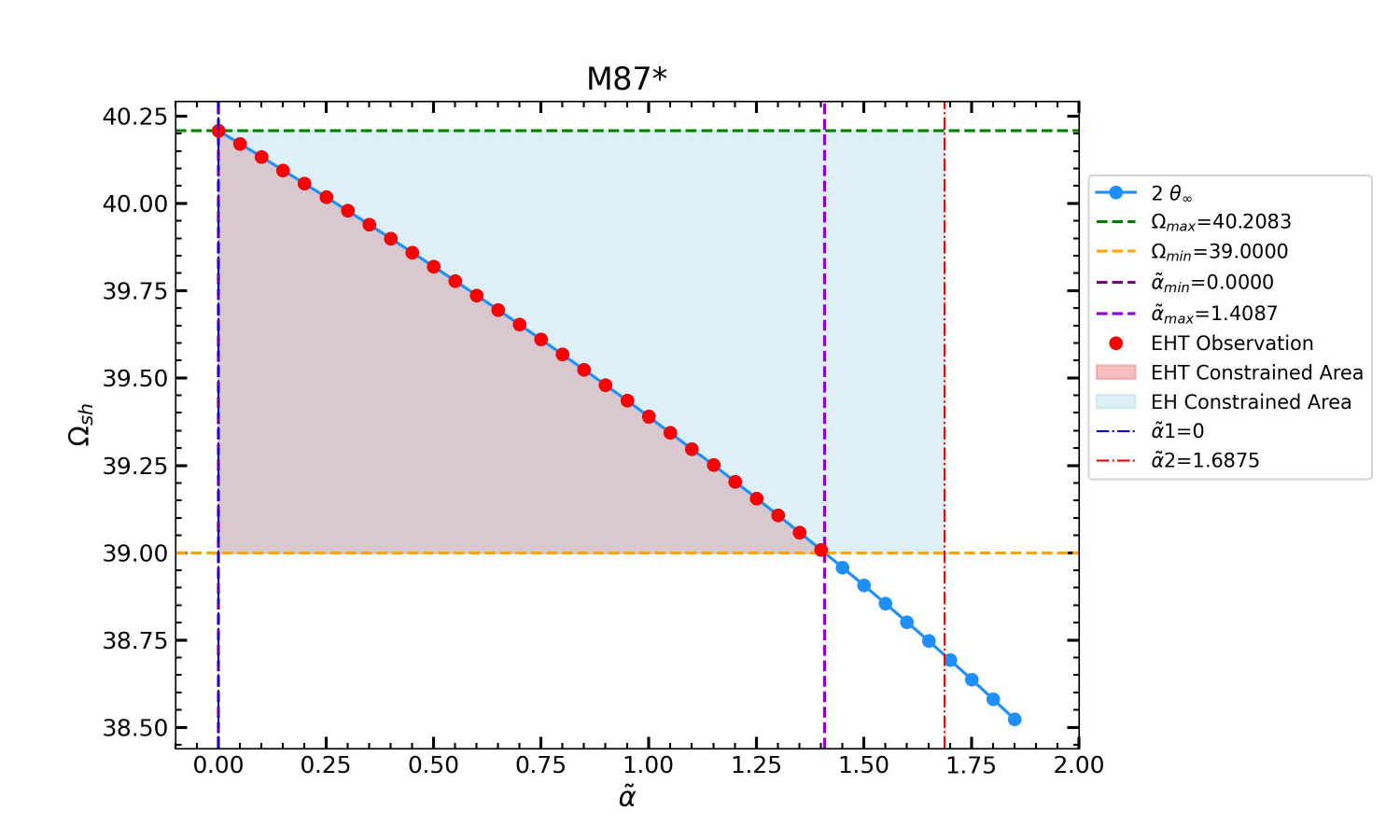}
\caption{
Using the $M87^*$ black hole as the research background, the blue dots represent the influence of the quantum correction parameter on the shadow ring when using a QCBH to simulate the supermassive $M87^*$ black hole. Here, the shadow ring diameter \(\Omega_{sh}\) is twice the angular position \(\theta_{\infty}\). The red dots represent the values within the first confidence interval for the $M87^*$ black hole shadow observed by the EHT. The blue area indicates the range of quantum correction parameter values for the existence of QCBH. The light red area represents the constraint range imposed by the EHT. The red dashed line marks the maximum value for the existence of an event horizon, and other corresponding dashed lines denote boundary values. The unit of the shadow ring diameter is $\mu as$.}
\label{g}
\end{figure}

For the supermassive black hole $M87^*$, in 2019, the EHT collaboration obtained the first-ever image of the supermassive black hole \(M87^*\). Their data indicated that the diameter of the black hole's ring structure (i.e., the shadow) is \(\Omega_{sh} = 2\theta = 42 \pm 3 \mu as \) \cite{EventHorizonTelescope:2019dse,EventHorizonTelescope:2019ggy}. Therefore, the next step is to apply the QCBH to the $M87^*$ black hole and constrain the quantum correction parameters using the observational data to ensure that the diameter of its ring structure falls within the first confidence interval \(\sigma\).
As shown in Figure \ref{g}, the confidence interval for the Event Horizon Telescope's observation of the M87* black hole shadow is represented by the light red area, while the blue area represents the constraint region of the black hole event horizon. Clearly, the intersection of these two regions indicates where the quantum correction parameter is constrained, i.e., the constrained range.
From the figure, it is easy to see that the range of values for the quantum correction parameter, constrained by the black hole event horizon and the EHT, is \(0 \leq \tilde{\alpha} \leq 1.4087\), which translates back to the original parameter space as \(0 \leq \frac{\alpha }{ M^2} \leq 1.4087\). Within this constraint, the QCBH always possesses event horizons and does not exceed its limit (the limit for the existence of event horizons is \(\tilde{\alpha} = 1.6875\)). This indicates that the QCBH can well match the shadow characteristics of astrophysical black holes, providing a basis for distinguishing QCBH from SBH in the near future.

\begin{figure}[]
\includegraphics[width=0.5 \textwidth]{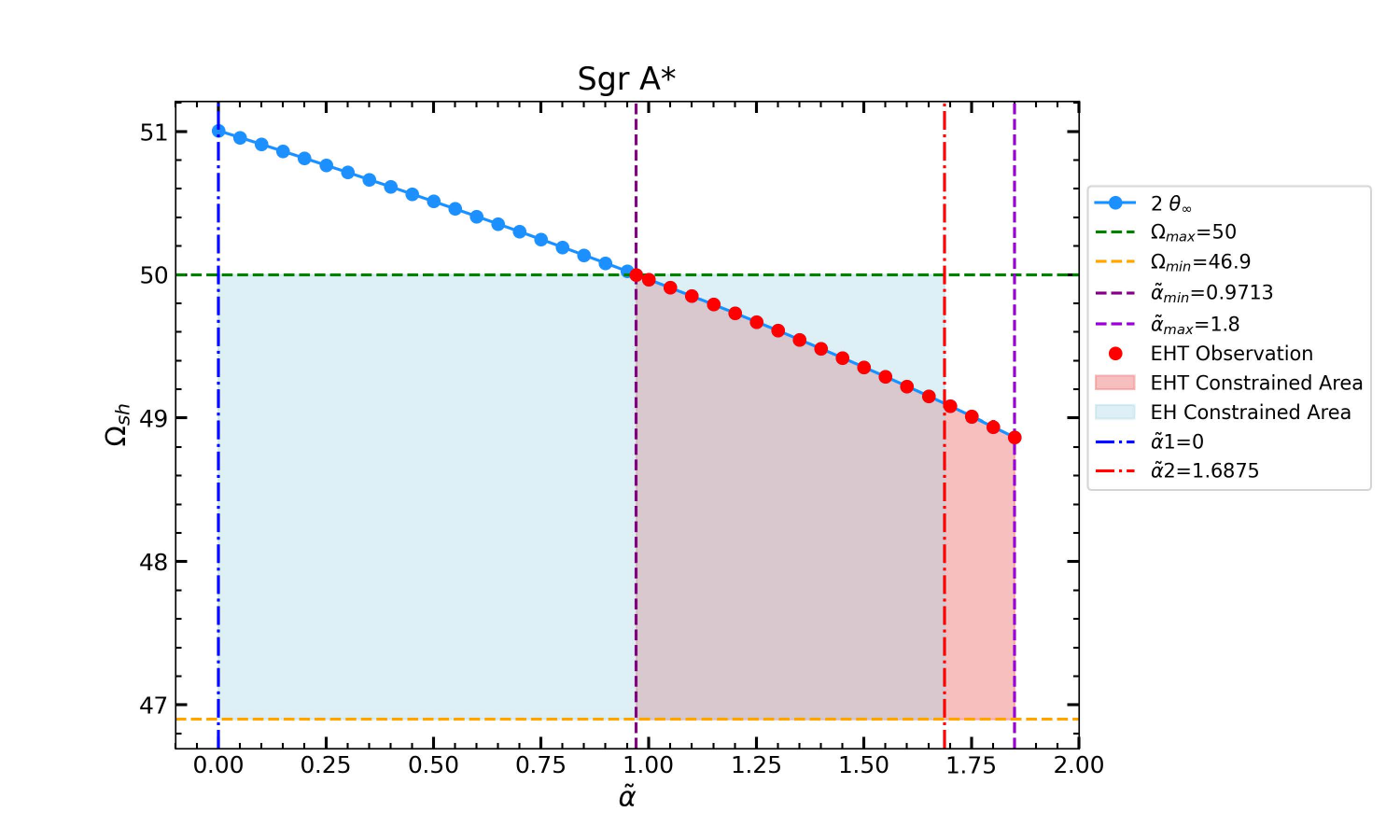}
\caption{
Using the $SgrA^*$ black hole as the research background, the blue dots represent the influence of the quantum correction parameter on the shadow ring when using a QCBH to simulate the supermassive $SgrA^*$ black hole. Here, the shadow ring diameter \(\Omega_{sh}\) is twice the angular position \(\theta_{\infty}\). The red dots represent the values within the first confidence interval for the $SgrA^*$ black hole shadow observed by the EHT. The blue area indicates the range of quantum correction parameter values for the existence of QCBH, which is the constrained region. The light red area represents the constraint range imposed by the EHT. The red dashed line marks the maximum value for the existence of an event horizon, and other corresponding dashed lines denote boundary values. The unit of the shadow ring diameter is $\mu as$.}
\label{h}
\end{figure}

For the supermassive $SgrA^*$ black hole, in 2022, the EHT team conducted observations of the $Sgr A^*$ black hole at the center of the Milky Way. In the literature \cite{EventHorizonTelescope:2022xqj}, they obtained the average shadow diameter of the supermassive $Sgr A^*$ black hole using three independent algorithms (eht-imaging, SMILI, and DIFMAP), with \(\Omega_{sh} \in (46.9, 50.0)\mu as\) and a $68\%$ confidence interval of \(\theta_{sh} \in (41.7, 55.6)\mu as\). Clearly, the quantum correction parameter is strongly constrained by the average shadow diameter \(\Omega_{sh}\).
As shown in Figure \ref{h}, similarly, the light red area represents the range of the $SgrA^*$ black hole shadow observed by the EHT that falls within the first confidence interval, while the blue area represents the constraint range for the event horizon of the QCBH. The intersection of these two areas indicates the constraint interval for the quantum correction parameter imposed by the EHT.
From the figure, it is easy to see that the range of values for the quantum correction parameter is constrained to \(0.9713 \leq \tilde{\alpha} \leq 1.6715\), which translates back to the original parameter space as \(0.9713 \leq \frac{\alpha }{ M^2} \leq 1.6715\). This means that if the value of the quantum correction parameter falls within this constrained range, the shadow of the QCBH will be consistent with the shadow of the $Sgr A^*$ black hole observed by the EHT.


\section{\label{sec:6}Discussion and conclusions}

Gravitational lensing provides an important window for exploring extreme celestial bodies and physical phenomena in the universe. In recent years, LQG theory has become a prominent area of research. The QCBH model introduces quantum effects to modify the structure and behavior of black holes beyond the framework of classical general relativity. These modifications not only resolve the singularity problem in classical black hole models but also potentially offer new predictions regarding the properties of event horizons, the evolution of black holes, and their radiation characteristics \cite{Lewandowski:2022zce,Ashtekar:2005qt,Modesto:2004xx,Haggard:2014rza,Gambini:2013exa}. These quantum effects may lead to changes in the horizon radius and the unstable photon sphere radius of QCBH, thereby affecting their gravitational lensing effects. This makes them an important avenue for exploring black hole properties and quantum gravity effects.

Based on these considerations, we investigated the impact of the quantum correction parameter on lensing coefficients and assumed a QCBH as a candidate for the supermassive black holes $M87^*$ and $Sgr A^*$. We explored the influence of the quantum correction parameter \(\tilde{\alpha}\) on image positions and the Einstein ring. Additionally, we used the EHT observations of the shadows of the supermassive black holes $M87^*$ and $Sgr A^*$ to constrain the value of the quantum correction parameter. Specifically, we studied how gravitational lensing images under QCBH change with varying quantum correction parameter \(\tilde{\alpha}\), including shifts in image positions, changes in the Einstein ring radius, and the constraint range of the quantum correction parameter. These findings will help us better understand the manifestation of quantum effects in actual astrophysical environments and provide guidance for future observations. The specific results are as follows:

In the strong-field limit, the gravitational lensing deflection angle and the corresponding coefficients for QCBH were calculated using the method of Bozza et al. Numerical computations show that the lensing coefficient \(\bar{a}\) increases with the quantum correction parameter \(\tilde{\alpha}\), while the deflection angle \(\alpha_D\) and the lensing coefficient \(\bar{b}\) decrease as \(\tilde{\alpha}\) increases. Furthermore, when the QCBH degenerates into a SBH, our results are \(\bar{a} = 1\) and \(\bar{b} = -0.40023\), which are in complete agreement with the SBH lensing coefficient values \cite{Bozza:2002zj}.

Using $M87^*$ and $Sgr A^*$ black holes as models for QCBH, we study their Einstein rings, relativistic images, and same-side time delays. The results indicate that the quantum correction parameter has a significant impact on the Einstein rings (see Figure \ref{e} and Table \ref{table1}). The deviation of the Einstein ring for $M87^*$ is below $0.7256 \mu as$, and for $Sgr A^*$ it is below $0.9205 \mu as$ (compared to the SBH).
As the quantum correction parameter increases, both the angular position $\theta_{\infty}$ of the relativistic images and the brightness ratio $r_{mag}$ between the images decrease, while the image separation $S$ increases. The angular position $\theta_{\infty}$ of the relativistic images ranges from $20.1042\mu as$ to $19.3535\mu as$ for $M87^*$, and from $25.5026\mu as$ to $24.5504\mu as$ for $SgrA^*$ as the quantum correction parameter varies. We also calculated the deviation and time delay between the QCBH and the SBH.
For the $M87^*$ black hole, the deviation in the angular position reaches \( |\delta(\theta_\infty)| = 0.7507 \, \mu as \), and the deviation in the image separation reaches \( |\delta(S)| = 0.025 \, \mu as \) (see Table \ref{table3}). The time delay can be as high as $293.9829$ hours, with the deviation in time delay between the QCBH and the SBH reaching $10.9766$ hours, which is sufficient for astronomical observation (see Tables \ref{table2} and \ref{table3}).
In the case of the $Sgr A^*$ black hole, simulating the QCBH, the deviation in the angular position reaches \( |\delta(\theta_\infty)| = 0.9522 \, \mu as \), and the deviation in the image separation reaches \( |\delta(S)| = 0.0317 \, \mu as \) (see Table \ref{table3}). The time delay reaches $10.8548$ minutes, with a relative deviation of $0.4393$ minutes, which is evidently too short for observational purposes (see Tables \ref{table2} and \ref{table3}).
In other words, these ranges are consistent with the existing observational range of the EHT for supermassive black holes. However, due to the current resolution of the EHT being approximately $20\mu as$, the existing equipment cannot accurately distinguish these differences. The next generation EHT is expected to resolve this issue. Once we can distinguish two relativistic images, we will be able to differentiate between the SBH and the QCBH, thereby further deepening our understanding of the properties of quantum correction parameter.

Based on the observational data from the EHT of the supermassive black holes $M87^*$ and $Sgr A^*$, we can effectively constrain the range of the quantum correction parameter. 
By analyzing the observed range of the black hole shadow diameter within the first confidence interval, the QCBH model shows results that are highly consistent with the actual observational data.
Specifically, with $M87^*$ as the study background, the quantum correction parameter is constrained within the range \(0 \leq \frac{\alpha }{ M^2} \leq 1.4087\), which completely avoids the scenario of no event horizon. In the case of $Sgr A^*$, the quantum correction parameter is constrained within the range \(0.9713 \leq \frac{\alpha }{ M^2} \leq 1.6715\). These results indicate that the QCBH model is not only theoretically reasonable but also shows a high degree of agreement with actual observations, providing a solid foundation and direction for future studies on the differences between QCBH and classical black holes.

In conclusion, the QCBH is not merely a theoretical construct but has the potential to become a viable candidate for astrophysical black holes.
This is because, on the one hand, the QCBH exhibits a high degree of consistency with actual astronomical observations.
On the other hand, the QCBH model avoids spacetime singularities, making it more aligned with the conditions of the real universe.
At the same time, our numerical simulations also indicate that the QCBH has reached the observational range of the current EHT (e.g., deviations in angular position on the order of $\sim 1 \mu as$, deviations in brightness ratio on the order of $\sim 0.01 \mu as$, time delay deviations of several tens of hours, and the black hole shadow highly matching observational data).
Unfortunately, due to the resolution limits of current equipment, it is currently impossible to distinguish these differences.
However, achieving this level of precision is only a matter of time, and the next generation of the EHT is expected to reach such precision.
Therefore, it is hoped that in the near future, we will be able to accurately distinguish between two relativistic images to explore the properties of QCBH and differentiate them from SBH.
Furthermore, if the rotational solution of the QCBH model can be found, exploring the lensing effects based on this will be very meaningful, marking a significant direction for future research.

\section{acknowledgements}
We acknowledge the anonymous referee for a constructive report that has significantly improved this paper. This work was supported by the Special Natural Science Fund of Guizhou University (Grant No.X2022133), the National Natural Science Foundation of China (Grant No. 12365008) and the Guizhou Provincial Basic Research Program (Natural Science) (Grant No. QianKeHeJiChu-ZK[2024]YiBan027) . 


\end{document}